\documentclass[letterpaper, 10 pt, conference]{ieeeconf}  

\IEEEoverridecommandlockouts                              

\usepackage{cite}
\usepackage{amsmath,amssymb,amsfonts}
\usepackage{graphicx} 
\usepackage{amsmath}
\usepackage{amssymb}  
\usepackage{mathrsfs}
\usepackage{multicol}
\usepackage{xcolor}
\usepackage{float}
\usepackage{subcaption}
\usepackage{mathtools}
\usepackage{soul}
\usepackage{bm}
\usepackage[T1]{fontenc}
\newtheorem{rem}{Remark}

\newtheorem{defn}{Definition}
\newtheorem{prop}{Proposition}
\newtheorem{prob}{Problem}

\newtheorem{theorem}{Theorem}

\usepackage{tikz}
\usepackage{xcolor}
\usepackage{dsfont}
\usepackage{algorithm}
\usepackage{algpseudocode}

\usepackage{hyperref} 

\definecolor{darkspringgreen}{rgb}{0.09, 0.45, 0.27}

\def\BibTeX{{\rm B\kern-.05em{\sc i\kern-.025em b}\kern-.08em
    T\kern-.1667em\lower.7ex\hbox{E}\kern-.125emX}}

\title{\LARGE \bf
Unconstrained learning of networked nonlinear systems via free parametrization of stable interconnected operators 
}

\author{Leonardo Massai, Danilo Saccani, Luca Furieri and Giancarlo Ferrari-Trecate
\thanks{This research has been supported by the Swiss National Science Foundation under the NCCR Automation (grant agreement 51NF40\_180545).}
\thanks{The authors are with the Institute of Mechanical Engineering, Ecole Polytechnique Fédérale de Lausanne (EPFL), CH-1015 Lausanne, Switzerland. (email: \tt\small {\{l.massai,danilo.saccani,luca.furieri, giancarlo.ferraritrecate\}@epfl.ch) } }%
}

\begin{document}

\maketitle
\thispagestyle{empty}
\pagestyle{empty}

\begin{abstract}

This paper characterizes a new parametrization of nonlinear networked incrementally $L_2$-bounded operators in discrete time. The distinctive novelty is that our parametrization is \emph{free} -- that is, a sparse large-scale operator with bounded incremental $L_2$ gain is obtained for any choice of the real values of our parameters. This property allows one to freely search over optimal parameters via unconstrained gradient descent, enabling direct applications in large-scale optimal control and system identification. Further,  we can embed prior knowledge about the interconnection topology and stability properties of the system directly into the large-scale distributed operator we design. 
Our approach is extremely general in that it can seamlessly encapsulate and interconnect state-of-the-art Neural Network (NN) parametrizations of stable dynamical systems. 
To demonstrate the effectiveness of this approach, we provide a simulation example showcasing the identification of a networked nonlinear system. The results underscore the superiority of our free parametrizations over standard NN-based identification methods where a prior over the system topology and local stability properties are not enforced.



\end{abstract}


\section{Introduction}
Interconnected systems are composed of multiple subsystems interacting over physical or cyber coupling channels and encompass diverse engineering applications like intelligent buildings, power grids, and transportation networks. Typically, effective control and monitoring of interconnected systems hinge on precise models of the subsystems. Failure to accurately and promptly identify the dynamics of a single subsystem can trigger network-wide instability, primarily because of uncertainty propagation through the coupling channels.
This renders conventional system identification methods tailored for single-agent systems unsuitable for this scenario~\cite{ljung1998system}. 

To improve on this matter, we want to exploit the prior knowledge about the model that we want to identify in terms of its structure and stability properties to define a class of distributed operators that capture such a structure while also ensuring some stability properties.
More in detail, in this work we focus on the stability property known as incremental $L_2$-boundedness, which is a particular case of incremental dissipativity.

When dealing with linear models, incremental $L_2$-boundedness~\cite{koelewijn2021incremental} can be enforced by imposing Linear Matrix Inequality (LMI) constraints. This approach has been extended to certain families of non-linear systems via the Linear Parameter Varying framework \cite{koelewijn2021incremental}. 
In system identification, this means that if we want to enforce incremental $L_2$-boundedness we end up with a constrained optimization problem. However, this is not desirable for several reasons such as poor scalability, feasibility issues, and computational efficiency.

With this in mind, in this paper, we introduce a new parametrization of distributed incremental $L_2$-bounded operators. The parameterization is free in the sense that for any value of the parameters, the resulting distributed operator is incrementally $L_2$-bounded. Consequently, this parametrized set of operators can be applied to system identification without the need to impose additional constraints for incremental $L_2$-boundedness, hence allowing one to solve an unconstrained optimization problem. 

\subsubsection*{Related works}
Dissipativity theory has found applications in a wide range of control problems~\cite{willems1972dissipative,byrnes1994losslessness}. It offers a unified framework for analyzing stability and performance in nonlinear systems~\cite{willems1972dissipative}. Several frameworks exist to analyze these properties ~\cite{rajpurohit2016dissipativity,verhoek2023convex,hill1976stability} and ensure global stability of nonlinear systems~\cite{angeli2002lyapunov,pavlov2008incremental}.
However, the dissipativity and stability properties of the subsystems can be preserved or altered when the systems are combined~\cite{moylan1978stability}. Different works have derived sufficient conditions to provide guarantees on input-output properties for generic interconnected systems~\cite{arcak2016networks,inoue2018dissipativity}.
This framework has recently been exploited to enforce robustness in Neural Networks (NNs) by analyzing them from a dynamical system perspective.
Recurrent Equilibrium Networks (RENs)~\cite{revay2023recurrent,revay2021recurrent,martinelli2023unconstrained} achieve incremental $L_2$-boundedness through a specific parameterization, enabling unconstrained training while guaranteeing stability. 
While NNs have found attractive applications in various fields, including the identification of dynamical systems~\cite{terzi2020learning,chen1990non,zakwan2022physically}, their full potential is not yet fully understood. While certain NN architectures can offer dissipative properties,  it remains a challenge to develop a parameterization method that ensures these properties are maintained when the network has a specific interconnection topology.
\subsubsection*{Contributions}
The main contribution of this paper is the development of a free parametrization for interconnected incremental $L_2$-bounded operators. Specifically, we derive an explicit mapping from a set of freely chosen parameters to a set of parametrized sub-operators such that their interconnection through a desired topology is also incrementally $L_2$-bounded. 
This result is fully compatible with any free parametrization of the single sub-operators, as long as these parametrizations ensure a finite incremental $L_2$-bound for each sub-operator as well. This allows for the exploitation of pre-existing state-of-the-art parameterizations of incrementally $L_2$-bounded sub-operators (such as RENs~\cite{revay2023recurrent}), thus opening the door to unconstrained optimization over distributed operators that also enjoy such a stability property. This is key as it enables the use of distributed NN operators in large-scale nonlinear system identification as well as control applications~\cite{furieri2022neural} while ensuring stability properties throughout the whole training phase.  

\section*{Notation}
Throughout the paper, we denote with $\mathbb{N}$ the set of non-negative integers.
$I_{a\times b}$ is the identity matrix with dimensions $a\times b$ and with 
$0_{a\times b}$ we denote a zero matrix of size $a\times b$, while $I_a$ and $0_a$ are square matrices with size $a$.
Positive semidefinite matrices $A$ are denoted as $A \succeq 0$ and $A_{[ij]}$ denotes the element of matrix $A$ corresponding to row $i$ and column $j$. We denote with blkdiag$(A,B)$ a block diagonal matrix created by aligning the matrices $A$ and $B$ along the diagonal and with col$_{j\in \mathcal{N}}(v_j)$, with $\mathcal{N} \subseteq \mathbb{N}$ a vector which consists of the stacked
subvectors $v_j$ where $j \in \mathcal{N}$  while $ v_{0: T}$ denotes a sequence of $v(t)$ with $t$ ranging from $0$ to $T$. The set of functions or sequences from $\mathbb{X}$ to $\mathbb{Y}$ is denoted by $\mathbb{Y}^{\mathbb{X}}$.
\section{Preliminaries}
The $L_2$-gain is a fundamental concept in the stability analysis and control of dynamical systems and it has been applied to ensure stability properties in networks of systems~\cite{dashkovskiy2007iss,arcak2016networks}. We leverage this concept to derive trainable incrementally $L_2$-bounded distributed operators. Now we recap some basic definitions regarding $L_2$-gain based on results presented in~\cite{koelewijn2021incremental,van2000l2}. \\
We consider parametrized discrete-time systems $\Sigma_{\theta} $ in the form:
\begin{equation} \label{eq:prel_statespace}
    \Sigma_{\theta}:     \begin{cases}
 x(t+1) &=f_{\theta}(x(t), u(t)), \quad x \in \mathbb{R}^n, u \in  \mathbb{R}^m \\
 y(t) &=h_{\theta}(x(t), u(t)), \quad y \in \mathbb{R}^p \\
 x(0)&=x_0, \quad \quad \quad \quad \quad \ \ t=0,\dots,T
\end{cases},
\end{equation}
where $x(t) \in \mathbb{R}^{n}, u(t) \in \mathbb{R}^{m}, y(t) \in \mathbb{R}^{p}, \theta \in \mathbb{R}^q $  are the system state, input, output and parameter respectively. $f_{\theta}: \mathbb{R}^q \times \mathbb{R}^{n} \times \mathbb{R}^{m} \mapsto \mathbb{R}^{n}$ and $h_{\theta}: \mathbb{R}^q \times \mathbb{R}^{n} \times \mathbb{R}^{m} \mapsto \mathbb{R}^{p}$ are, instead, assumed to be Lipschitz continuous functions depending on $\theta$, such that $f_{\theta}(0,0)=0$ and $h_{\theta}(0,0)=0$ for all $\theta$, and such that for all initial conditions $x_0\in\mathbb{R}^n$ there is a unique solution $(x,u,y)\in (\mathbb{R}^{n}\times \mathbb{R}^{m}\times \mathbb{R}^{p} )^{\mathbb{N}}$. We define the set of solutions of~\eqref{eq:prel_statespace} as $$\mathcal{B}\coloneq \{ (x,u,y)\in (\mathbb{R}^{n}\times \mathbb{R}^{m}\times \mathbb{R}^{p} )^{\mathbb{N}} | (x,u,y) \text{ satisfies}~\eqref{eq:prel_statespace} \}.$$
Notice that the system $\Sigma_{\theta}$ given by~\eqref{eq:prel_statespace} can be seen as a parametrized operator that, for a fixed value of the parameter, takes as input an initial condition $x(0) \in \mathbb{R}^{n_i}$ as well as a sequence $u_{0:T}$ and outputs a sequence $y_{0:T}$ for any given $T \in \mathbb{N}$. In this sense, we will also use the notation $    y_{0:T} = \Sigma_{\theta}(x_0, u_{0:T})$ to highlight the mapping operated by $\Sigma_{\theta}$.
Notably, recursive operators, such as~\eqref{eq:prel_statespace}, can be viewed as deep neural networks unrolled in time. Consequently, these operators can serve as operators in learning problems and be optimized through Backpropagation Through Time (BPTT)~\cite{furieri2022neural}.
We can provide the following definitions.
\begin{defn}{($L_2$-gain)~\cite{byrnes1994losslessness}}
    The operator $\Sigma_{\theta}$ given by~\eqref{eq:prel_statespace} has a finite $L_2$-gain$\leq$ $\gamma$ if it is dissipative with respect to the supply rate $s(u,y)=\gamma^2 \| u \|^2 - \| y \|^2$; that is, there exists a storage function $V:\mathbb{R}^n\rightarrow \mathbb{R}^+$ such that for all $t\in\mathbb{N}$
    \begin{equation}
        V(x(t+1)) - V(x(t)) \leq s(u(t),y(t)),
    \end{equation}
\end{defn}
The storage function $V$ can be interpreted as a model of the stored ``energy'' in the system w.r.t. a single point of neutral storage (minimum energy).
\begin{defn}{(Incremental $L_{i2}$ gain~\cite{koelewijn2021incremental})}
\label{def:li2}
    The operator $\Sigma_{\theta}$ in the form~\eqref{eq:prel_statespace} is said to have a finite incremental $L_2$-gain $\gamma$, denoted as $L_{i2}$-gain, if for all $t\in\mathbb{N}$ and $(x,u,y),(\hat{x},\hat{u},\hat{y})\in\mathcal{B}$ it is dissipative with respect to the supply rate $s(u,\hat{u},y,\hat{y})=\gamma^2 \| u-\hat{u} \|^2 - \| y-\hat{y} \|^2$; that is, there exists an \textit{incremental storage function} $V:\mathbb{R}^n\times \mathbb{R}^n\rightarrow \mathbb{R}^+$ with $V(0,0)=0$ such that for all $t\in\mathbb{N}$:
    \begin{multline}
        V(x(t+1),\hat{x}(t+1)) - V(x(t),\hat{x}(t)) \leq \\ s\left(u(t),\hat{u}(t),y(t),\hat{y}(t)\right).
    \end{multline}
\end{defn}
We can now introduce the following set of parameters $\theta$:
\begin{align}
    \Phi_{\gamma} = \{ \theta \in \mathbb{R}^q \: | \:  \Sigma_{\theta} \mbox{ has finite $L_{i2}$-gain } \gamma \}. 
\end{align}
In a learning problem, such as in system identification, we aim to fit a set of data by tuning the parameters of an operator $\Sigma_{\theta}$ by minimizing a certain, differentiable, loss function. However, it's often desirable to also preserve the stability properties of the underlying system, obtaining a constrained optimization problem. By considering a constraint on the incremental $L_2$-boundedness of the system, we can write the identification problem as follows:
\begin{subequations}
 \label{eq:optc}
 \begin{align}
  \label{eq:optc1}
         \displaystyle\min_{\theta} \quad  &\mathcal{L} \left( y_{0:T}, \tilde{y}_{0:T} \right) \nonumber\\[2ex]
         \mbox{s.t.} \quad & y^i_{0:T} = \Sigma_{\theta} \left(\tilde{x}^i_0, \tilde{u}^i_{0:T} \right), \quad i=1,\dots n_s  \\[2ex]
          \label{eq:optc2}
         & \theta \in \Phi_{\gamma},
 \end{align}
\end{subequations}

\medskip
\noindent where $(\tilde{u}^i_{0:T},\tilde{y}^i_{0:T},\tilde{x}^i_0)$ denote finite sequences of input-output measurements and the initial condition, and $n_s$ is the number of trajectories considered.
For instance, for system identification purposes, one might want to minimize the sum of the squared residuals over all the sequences, i.e.,
    \begin{equation} \label{eq:loss}
        \mathcal{L} \left( y_{0:T}, \tilde{y}_{0:T} \right) = \sum_{i=1}^{n_s} \| y^i_{0:T} - \tilde{y}^i_{0:T} \|^2
    \end{equation}
\medskip
The problem~\eqref{eq:optc} is a constrained optimization problem that, when $\Sigma_{\theta}$ is a linear operator, features linear convex constraints \eqref{eq:optc2} that can be written as LMIs~\cite{chesi2021exact}. This is not desirable because of poor computational efficiency as well as scalability and feasibility issues. Notice that when $\Sigma_{\theta}$ is a non-linear operator, all these issues are exacerbated even further, as the general problem tends to become non-convex and non-linear even in the constraints.

For these reasons, one would like to cast the problem \eqref{eq:optc} as an unconstrained optimization problem by getting rid of \eqref{eq:optc2}, even at the cost of restricting the set of feasible parameters. This is key as it will open the door for the utilization of parametrized families of highly non-linear operators in the form \eqref{eq:prel_statespace}, such as neural networks, which can be optimized very efficiently by solving an unconstrained optimization problem with standard backpropagation and unconstrained gradient descent methods. 
All of this can be obtained by freely parametrizing the set $\Phi_{\gamma}$. 

\smallskip

\begin{defn}
    Given a non-empty set $\Psi \subseteq \mathbb{R}^a$, a parametrization of $\Psi$ is a mapping $\psi : \Theta \subseteq \mathbb{R}^q \mapsto \Psi $. the parametrization $\psi$ is called free if $\Theta = \mathbb{R}^q$.
\end{defn}

\smallskip

We will consider a free parametrization of the set $\Phi_{\gamma} $, that is, a mapping $\psi(\xi, \gamma) : \mathbb{R}^q  \mapsto  \Phi_{\gamma}  $ where $\xi$ are the free parameters. In other words, $\theta = {\psi(\xi, \gamma)} \in \Phi_{\gamma}, \: \forall \xi \in \mathbb{R}^q$ or, equivalently, $\Sigma_{\psi(\xi, \gamma)}$ has $L_{i2}$-gain $\gamma$, $\forall \xi \in \mathbb{R}^q$. Notice that here $\gamma$ is fixed but could also be treated as an additional free parameter of $\psi$ through a proper mapping such as $\gamma = z^2$ where $z \in \mathbb{R}$ is the new parameter.

Notice that if one can exhibit such a free parametrization, then that can be used to reformulate \eqref{eq:optc} as an unconstrained problem by optimizing over the new parameter $\xi$ as the constraint \eqref{eq:optc2} is satisfied for any value of $\xi$ by definition. 

\smallskip

\begin{rem}
One remarkable example of such free parameterization is given by operators called Recurrent Equilibrium Networks (RENs) in~\cite{revay2021recurrent}, which result from the closed-loop interconnection of a discrete-time linear dynamical system with a static non-linearity. RENs extend various non-linear operators, including Recurrent Neural Networks (RNNs) and standard feed-forward neural networks. A significant contribution of~\cite{revay2021recurrent} lies in presenting a free parametrization for a class of RENs with stability and dissipativity properties, such as $L_{i2}$-boundedness. Because of such features, REN operators can be efficiently optimized in a learning problem using unconstrained techniques.

\begin{rem}
\label{rmk:cons}
    There is usually a ``price to pay'' for transforming the problem \eqref{eq:optc} into an unconstrained one via a free parametrization. This stems from the fact that, usually, the map $\psi$ is not surjective and it effectively describes only a subset of $\Phi_{\gamma}$, which in turn means that we are restricting ourselves to optimize over a subset of $L_{i2}$-bounded operators. This is also true for REN operators. However, as it is shown in~\cite{revay2021recurrent}, the advantages of having a free parametrization greatly outweigh such a drawback.
\end{rem}

\medskip

Notice that so far we have considered only one single centralized operator $\Sigma_{\theta}$ but, as mentioned in the introduction, we would like to use a distributed operator to better capture the interconnectedness of real systems and we want to do that while preserving the $L_{i2}$-boundedness property.  The free parametrization approach outlined earlier has never been generalized to distributed operators built by several interconnected sub-operators. It is worth mentioning that simply connecting $L_{i2}$-bounded sub-operators will not work, in general, as the resulting operator could fail to be $L_{i2}$-bounded. \\ 
We introduce a free parametrization for distributed operators in the next Section.

\end{rem}

\smallskip


\section{Setting, Models, and Problem formulation}

\begin{figure}
    \centering
\tikzset{every picture/.style={line width=0.75pt}} 

\begin{tikzpicture}[x=0.75pt,y=0.75pt,yscale=-1,xscale=1]

\draw   (241,41) -- (320.6,41) -- (320.6,120) -- (241,120) -- cycle ;
\draw   (260.57,130.43) -- (300.57,130.43) -- (300.57,169.43) -- (260.57,169.43) -- cycle ;
\draw    (321.4,80) -- (389,80) ;
\draw [shift={(392,80)}, rotate = 180] [fill={rgb, 255:red, 0; green, 0; blue, 0 }  ][line width=0.08]  [draw opacity=0] (10.72,-5.15) -- (0,0) -- (10.72,5.15) -- (7.12,0) -- cycle    ;
\draw    (340.71,149.86) -- (340.31,80.4) ;
\draw    (340.71,149.86) -- (303.77,149.83) ;
\draw [shift={(300.77,149.83)}, rotate = 0.04] [fill={rgb, 255:red, 0; green, 0; blue, 0 }  ][line width=0.08]  [draw opacity=0] (10.72,-5.15) -- (0,0) -- (10.72,5.15) -- (7.12,0) -- cycle    ;
\draw    (205.24,80.08) -- (236,80.01) ;
\draw [shift={(239,80)}, rotate = 179.87] [fill={rgb, 255:red, 0; green, 0; blue, 0 }  ][line width=0.08]  [draw opacity=0] (10.72,-5.15) -- (0,0) -- (10.72,5.15) -- (7.12,0) -- cycle    ;
\draw   (194.51,80.08) .. controls (194.51,77.12) and (196.92,74.71) .. (199.88,74.71) .. controls (202.84,74.71) and (205.24,77.12) .. (205.24,80.08) .. controls (205.24,83.04) and (202.84,85.44) .. (199.88,85.44) .. controls (196.92,85.44) and (194.51,83.04) .. (194.51,80.08) -- cycle ;
\draw    (200.43,150.43) -- (200.58,88.87) ;
\draw [shift={(200.59,86.87)}, rotate = 90.14] [color={rgb, 255:red, 0; green, 0; blue, 0 }  ][line width=0.75]    (10.93,-3.29) .. controls (6.95,-1.4) and (3.31,-0.3) .. (0,0) .. controls (3.31,0.3) and (6.95,1.4) .. (10.93,3.29)   ;
\draw    (260.26,150.11) -- (200.43,150.43) ;
\draw    (146,80) -- (191.51,80.07) ;
\draw [shift={(194.51,80.08)}, rotate = 180.09] [fill={rgb, 255:red, 0; green, 0; blue, 0 }  ][line width=0.08]  [draw opacity=0] (10.72,-5.15) -- (0,0) -- (10.72,5.15) -- (7.12,0) -- cycle    ;
\draw  [dash pattern={on 4.5pt off 4.5pt}] (171,31) -- (365,31) -- (365,181) -- (171,181) -- cycle ;

\draw (247.6,46.6) node [anchor=north west][inner sep=0.75pt]    {$\Sigma _{\theta _{1}}$};
\draw (288.6,93.4) node [anchor=north west][inner sep=0.75pt]    {$\Sigma _{\theta _{N}}$};
\draw (269.2,70.4) node [anchor=north west][inner sep=0.75pt]    {$\ddots $};
\draw (271.77,142.83) node [anchor=north west][inner sep=0.75pt]    {$M$};
\draw (346,124.4) node [anchor=north west][inner sep=0.75pt]    {$y$};
\draw (213.29,60.17) node [anchor=north west][inner sep=0.75pt]    {$u$};
\draw (372.4,56) node [anchor=north west][inner sep=0.75pt]    {$e$};
\draw (150.8,56.57) node [anchor=north west][inner sep=0.75pt]    {$d$};
\draw (124.6,153.6) node [anchor=north west][inner sep=0.75pt]    {${\textstyle \Sigma _{\theta }^{M}}$};
\end{tikzpicture}
\caption{Interconnection of $N$ operators $\Sigma_{\theta_i}$.}
    \label{fig:interconnection}
\end{figure}
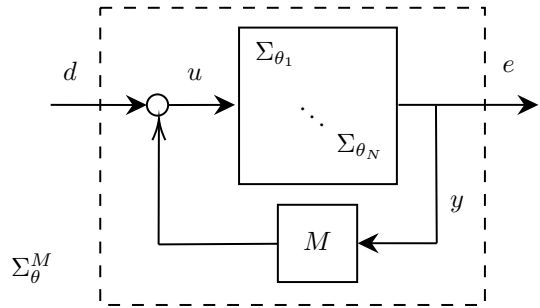
We consider $N$ parametrized sub-operators indexed by $i\in \mathcal{N}$  with $\mathcal{N}=\{1, \dots, N\}$
\begin{equation}
\label{eq:sys} \Sigma_{\theta_i} :
    \begin{cases}
               x_i(t+1) &  = f_{\theta_i}(x_i(t), u_i(t)),\\
       y_i(t) & =  h_{\theta_i}(x_i(t),u_i(t)),\quad  t=0,\dots,T
       \\
 x_i(0)&=x_{0i} 
    \end{cases},
\end{equation}
where $x_i(t) \in \mathbb{R}^{n_i}, u_i(t) \in \mathbb{R}^{m_i}, y_i(t) \in \mathbb{R}^{p_i}, \theta \in \mathbb{R}^q $ are the system state, input, output and parameter respectively.\\
These operators are interconnected, as shown in Fig.~\ref{fig:interconnection}\footnote{Notice that this is a special case of the framework introduced in~\cite{arcak2016networks}.}, to accommodate an
exogenous input $d\in\mathbb{R}^{m}$ and performance output $e\in\mathbb{R}^{p}$. More in detail, the interconnection topology is described by the matrix $\bar{M}$ that relates the sub-operators inputs and performance outputs as:
\begin{equation} \label{eq:interconnUE}
    \left[\begin{array}{l}
u \\
e
\end{array}\right]=\bar{M}\left[\begin{array}{l}
y \\
d
\end{array}\right]=\left[\begin{array}{ll}
M & I_{m} \\
I_{p} & 0_{m}
\end{array}\right]\left[\begin{array}{l}
y \\
d
\end{array}\right],
\end{equation}
where $u = \text{col}_{\mathcal{N}}(u_i)\in \mathbb{R}^m$, $y= \text{col}_{\mathcal{N}}(y_i)\in\mathbb{R}^p$. The matrix $\bar{M}$ describes how each sub-operator input $u_i$ is coupled with each sub-operator output $y_i$ as well as with the exogenous input $d_i$. More in detail, by developing~\eqref{eq:interconnUE} we get~$u=My+d$ and~$e=y$. The first equation states that the input $u$ of each sub-operator is a fully generic linear combination of the sub-operator's output $y$ via the matrix $M$ plus the exogenous input $d$, which is an additive to each input. As we indicated before, this choice constrains $d$ to have the same dimension of $u$, which is a hypothesis that we aim to relax in future work by considering a more general form for the matrix $\bar{M}$. 

The second equation $e=y$ describes the fact that the performance output of the interconnected operator coincides with all the outputs of each sub-operator. 

The coupling of each sub-operator $\Sigma_{\theta_i}$ via~\eqref{eq:interconnUE} produces an operator $\Sigma_{\theta}^{M} $ that, with a slight abuse of notation, can be written as
\begin{equation}
\label{eq:sysM} \Sigma_{\theta}^{M}:
    \begin{cases}
               x(t+1) &  = f_{\theta}\left(x(t), My(t)+d(t)\right)\\
       y(t) & =  h_{\theta}\left(x(t), M y(t)+d(t)\right)
       \\
 x(0)&=x_0,\quad \quad \quad t=0,\dots,T 
    \end{cases},
\end{equation}
where $x(t)=\text{col}_{i \in \mathcal{N}}(x_i(t)) \in \mathbb{R}^{n}, \theta=\text{col}_{i \in \mathcal{N}}(\theta_i) \in \mathbb{R}^{q}$ and the functions $f_{\theta}$ and $h_{\theta}$ are the stacked sub-functions $f_{\theta_i}$ and $h_{\theta_i}$ applied block-wise. We assume that the coupling, and therefore the operator $\Sigma_{\theta}^{M}$, is well-defined, i.e., the equation $y(t)  =  h_{\theta}\left(x(t), M y(t)+d(t)\right)$ admits a unique solution $y(t)$, given $x(t)$ and $d(t)$, for $t=0,\dots,T $.
Now we are ready to state the problem we aim to solve:
\begin{prob}
    Consider the operator $\Sigma_{\theta}^{M}$ as defined in~\eqref{eq:sysM}. We aim to find a free parametrization of the set 
\begin{align}
    \Phi_{\gamma_{M}} = \{ \theta_i, i \in  \mathcal{N} \: | \:  \Sigma_{\theta}^{M} \text{ has finite $L_{i2}$-gain } \gamma_{M} \}. 
\end{align}
    In other words, we want to find some mapping from a set of free parameters to a set of parameters for each sub-operator such that their coupling $\Sigma_{\theta}^{M}$ has finite $L_{i2}$-gain. 
    This enables free tuning of sub-operators while preserving $L_{i2}$-gain for the interconnected operator, hence allowing one to solve problem~\eqref{eq:optc} with any unconstrained optimization algorithm.
    
\end{prob}

\section{Main results}
To tackle the problem, first of all, we consider sub-operators verifying $\Sigma_{\theta_i} \in  \Phi_{\gamma_i}, i \in \mathcal{N}$ and such that they are dissipative with a positive definite, continuously differentiable incremental storage function $V_i(\cdot,\cdot) $ and the quadratic supply rate:
\begin{multline}
    s_i(u_i,\hat{u}_i,y_i,\hat{y}_i) = \\ \left[\begin{array}{l}
        u_i - \hat{u}_i\\
         y_i - \hat{y}_i
    \end{array}\right]^{\top}
\underbrace{\left[\begin{array}{ll}
\gamma_i^2 I_{m_i} & 0_{m_i\times p_i}\\
0_{p_i\times m_i} & -I_{p_i}
\end{array}\right]  }_{X_i}  
        \left[\begin{array}{l}
        u_i - \hat{u}_i\\
         y_i - \hat{y}_i
    \end{array}\right].
\end{multline}
The goal is now to certify the validity of a prescribed $L_{i2}$ gain $\gamma_{M}$ for the interconnected operator $\Sigma_{\theta}^{M}$, or in other words, its dissipativity with respect to the quadratic supply rate:
\begin{equation} \label{eq:performanceSupply}
    \left[\begin{array}{l}
d-\hat{d} \\
e-\hat{e}
\end{array}\right]^\top 
\underbrace{\left[\begin{array}{cc}
\gamma_{M}^2 I_m & 0_{m \times p} \\
0_{p \times m} & -I_p
\end{array}\right]}_W
\left[\begin{array}{l}
d-\hat{d} \\
e-\hat{e}
\end{array}\right].
\end{equation}
Now, let us consider the following candidate incremental storage function, defined as the weighted sum of the local storage functions
\begin{equation} \label{eq:overalStorage}
    V(x,\hat{x})=\alpha_1 V_1\left(x_1,\hat{x}_1\right)+\cdots+\alpha_N V_N\left(x_N,\hat{x}_N\right),
\end{equation}
with $\alpha_i > 0$. 
Thus, following the results in~\cite{arcak2016networks}, we want to satisfy the following inequality
\begin{multline} \label{eq:dissipativityInterconnection}
    \sum_{i=1}^N \alpha_i \left[ V_i(x_i(t+1),\hat{x}_i(t+1))- V_i(x_i(t),\hat{x}_i(t)) \right] \leq \\ \sum_{i=1}^N \alpha_i \left[\begin{array}{l}
        u_i -\hat{u}_i\\
         y_i -\hat{y}_i
    \end{array}\right]^{\top}
X_i
        \left[\begin{array}{l}
        u_i -\hat{u}_i\\
         y_i-\hat{y}_i 
    \end{array}\right].
\end{multline}
By rewriting the right-hand side of~\eqref{eq:dissipativityInterconnection}, we obtain:
\begin{equation} \label{eq:rightInterconn}
    \left[\begin{array}{l}
        u_1 -\hat{u}_1\\
        \quad \ \ \vdots \\
        u_N -\hat{u}_N\\
         y_1 -\hat{y}_1 \\
         \quad \ \ \vdots \\
         y_N -\hat{y}_N
    \end{array}\right]^{\top}
\bm{X}(\alpha_iX_i)
        \left[\begin{array}{l}
        u_1 -\hat{u}_1\\
        \quad \ \ \vdots \\
        u_N -\hat{u}_N\\
         y_1 -\hat{y}_1 \\
         \quad \ \ \vdots \\
         y_N -\hat{y}_N
    \end{array}\right],
\end{equation}
where 
\begin{equation} \label{eq:Xpi}
     \bm{X}(\alpha_iX_i)= \left[\begin{array}{ll} \Gamma_{N,m} & 0\\
0 & -A_{N,p}
\end{array}\right],
\end{equation}
with $\Gamma_{N,m}=\text{blkdiag}\{\alpha_1\gamma_1^2 I_{m_1},\dots,\alpha_N\gamma_N^2 I_{m_N}\}$ and  $A_{N,p}=\text{blkdiag} \{ \alpha_1I_{p_1},\dots,\alpha_NI_{p_N} \}$. \\
The storage function~\eqref{eq:overalStorage}, serves as a storage function for the interconnection and, if the right-hand side of~\eqref{eq:dissipativityInterconnection} is negative semidefinite in $y$, when $u$ is eliminated by considering $u=My$, then the interconnection is Lyapunov stable~\cite[Thm. 3.1]{arcak2016networks}.
The right-hand side of~\eqref{eq:dissipativityInterconnection},
rewritten as in~\eqref{eq:rightInterconn}, is dominated by the performance supply rate~\eqref{eq:performanceSupply} if 
\begin{equation} \label{eq:LMI1}
    \left[\begin{array}{l}
        u -\hat{u}\\
         y-\hat{y} \\
         d-\hat{d} \\
         e-\hat{e} 
    \end{array}\right]^{\top}
\left[\begin{array}{cc}
\bm{X}(\alpha_iX_i) & 0 \\
0 & -W
\end{array}\right]
        \left[\begin{array}{l}
        u -\hat{u}\\
         y-\hat{y} \\
         d-\hat{d} \\
         e-\hat{e} 
    \end{array}\right] \preceq 0,
\end{equation}
when the variables $u$ and $e$ are constrained by~\eqref{eq:interconnUE}. \\Substituting 
\begin{equation} \label{eq:interconnection}
    \left[\begin{array}{l}
        u\\
         y \\
         d \\
         e 
    \end{array}\right] = \left[\begin{array}{ll}
        M & I_{m}\\
         I_{p} & 0_{p\times m} \\
         0_{m\times p}&I_m \\
         I_p & 0_{p\times m} 
    \end{array}\right] \left[\begin{array}{l}
        y\\
         d \\
    \end{array}\right],
\end{equation}
in~\eqref{eq:LMI1} we obtain the following performance condition~\cite{arcak2016networks}:
 \begin{equation} \label{eq:conditionL2}
\left[\begin{array}{ll}
    M & I\\
     I & 0 \\
     0&I \\
     I & 0 
\end{array}\right]^\top
\left[\begin{array}{cc}
\bm{X}(\alpha_iX_i) & 0 \\
0 & -W
\end{array}\right]
\left[\begin{array}{ll}
    M & I\\
     I & 0 \\
     0&I \\
     I & 0 
\end{array}\right] \preceq 0
\end{equation}
where $\bm{X}(\alpha_iX_i)$ is as defined in~\eqref{eq:Xpi}.
For the sake of simplicity, from now on we omit the size of identity and zero matrices.
The previous computations prove the following proposition adapted from~\cite{arcak2016networks}.
\begin{prop}\label{prop:L2gain}
    If there exist $\alpha_i>0$, $i \in  \mathcal{N}$, such that~\eqref{eq:conditionL2} holds, then the operator $\Sigma_{\theta}^{M}$ has a finite incremental $L_{i2}$-gain $\gamma_{M}$.
\end{prop}
Now we have all the tools needed to derive our main result. 

\begin{theorem} \label{thm:freeparam}
The set $\Phi_{\gamma_{M}}$ is non-empty and admits a free parametrization such that each sub-operator $\Sigma_{\theta_i}$ is parametrized by $\theta_i = \psi_i\left( \xi_i, \nu(z_i) \right)$ with $i \in  \mathcal{N}$ and with free parameters, $\xi_i \in \mathbb{R}^{q_i}, z_i \in \mathbb{R}$ where 
$\nu(z_i) : \mathbb{R} \mapsto \mathbb{R}_+$ is given by:
\begin{align}
\label{eq:param}
    \nu(z_i)  = & \sqrt{\frac{1}{1+ \max_{j\in\ \mathcal{N}_{p_i}} \left(\sum_{k}|M_{[kj]}|\right) + {z_i}^2}} \notag \\
    & \cdot  \sqrt{\frac{\gamma_{M}^2}{\max_{j\in\ \mathcal{N}_{u_i}} \left(\sum_{k}|M_{[jk]}|\right) \gamma_{M}^2 +1}},
\end{align}
where $M_{[ij]}$ is the element in the $i$-th row and $j$-th column in matrix $M$ and 
\begin{align}
      &  \mathcal{N}_{u_i} = \mbox{set of indices of $u$ associated with $\Sigma_{\theta_i}, i \in \mathcal{N}$}, \notag \\
      &  \mathcal{N}_{y_i} = \mbox{set of indices of $y$ associated with $\Sigma_{\theta_i}, i \in \mathcal{N}$}.
\end{align}

\end{theorem}
\begin{proof}
By Proposition~\ref{prop:L2gain} the existence of a finite  $L_{i2}$-gain  $\gamma_{M}$ for $\Sigma_{\theta}^{M}$ depends on the inequality \eqref{eq:conditionL2} being satisfied.  

 In particular,  the condition~\eqref{eq:conditionL2} reads as 
\begin{equation}
\left[\begin{array}{ll}
-M^{\top}\Gamma_{N,m} M +A_{N,p}-I & -M^{\top} \Gamma_{N,m}\\[2ex]
-\Gamma_{N,m}M & - \Gamma_{N,m} +\gamma_M^2 I
\end{array}\right] \succeq 0  
\end{equation}
By applying the Schur complement we obtain the following conditions:
\begin{align} \label{eq:step1proof}
   & -M^{\top}\Gamma_{N,m} M +A_{N,p}-I  \notag\\
   & -M^{\top} \Gamma_{N,m}\left( - \Gamma_{N,m} +\gamma_M^2 I\right)^{-1} \Gamma_{N,m}M \succeq 0 \\[2ex]
   & - \Gamma_{N,m} +\gamma_M^2 I \succeq 0 \iff \alpha_i \gamma_i^2 \le \gamma_M^2, \: \forall i \in \mathcal{N} \label{eq:step1proof2}
   .
\end{align}
By collecting $M$ we can rewrite~\eqref{eq:step1proof} as
\begin{multline}
    M^{\top} ( -\Gamma_{N,m}+\Gamma_{N,m}\left(  \Gamma_{N,m} -\gamma_M^2 I\right)^{-1} 
    \Gamma_{N,m} )M\\+A_{N,p}-I \succeq 0.
\end{multline}
By applying the Schur complement again this is equivalent to:
\begin{align} \label{eq:SchurLMI}
    \left[\begin{array}{ll}
A_{N,p}-I & M^{\top} \\[2ex]
M & -D^{-1}
\end{array}\right] & \succeq 0, \\
D & \preceq 0 \label{eq:shur2}
\end{align}
where 
$    D = -\Gamma_{N,m}+\Gamma_{N,m}\left(  \Gamma_{N,m} -\gamma_M^2 I\right)^{-1} 
    \Gamma_{N,m}$.
Notice that $D$ is a diagonal matrix with $D_{ii}= \frac{\alpha_i \gamma_i^2  \gamma_M^2}{\alpha_i \gamma_i^2 -\gamma_M^2}$ and that \eqref{eq:shur2} is equivalent to \eqref{eq:step1proof2}.

Now we use Gershgorin theorem~\cite{salas1999gershgorin} to guarantee that~\eqref{eq:SchurLMI} is positive semi-definite. In particular, let us focus on the first $p$ rows of the matrix~\eqref{eq:SchurLMI}

We want all the Gershgorin circles to lie in the non-negative half-plane, i.e.,
\begin{align}
\label{eq:paramproof}
    \underbrace{\left(A_{N,p}-I\right)_{[jj]}}_{\mbox{Center}} -\underbrace{\sum_{k}|M_{[kj]}|}_{\mbox{Radius}} \ge 0, \quad  j \in \mathcal{N}_{u_i}
\end{align}

Notice that the Gershgorin center is $\alpha_i-1$ for every $j \in \mathcal{N}_{u_i}$, hence condition \eqref{eq:paramproof} can be enforced by imposing $\alpha_i$ such that:
\begin{align}
\label{eq:pesi}
    \alpha_i &= 1+ \max_{j\in\mathcal{N}_{p_i}} \left(\sum_{k}|M_{[kj]}|\right) + z_i^2 , \ \ \ i \in \mathcal{N}
\end{align}
where $z_i \in \mathbb{R}$. Similarly, we can impose the other Gershgorin circles associated with the remaining rows of the matrix~\eqref{eq:SchurLMI} to lie in the non-negative half of the plane, yielding the condition:

\begin{align}
\label{eq:paramproof2}
    -\frac{\alpha_i \gamma_i^2 - \gamma_{M}^2}{\alpha_i \gamma_i^2 \gamma_{M}^2} 
    - \max_{j\in\mathcal{N}_{u_i}} \left(\sum_{k}|M_{[jk]}|\right)
     \geq 0 , \ \ \ i \in \mathcal{N}\\
    0 \leq \alpha_i \gamma_i^2 \leq \frac{\gamma_M^2}{\max_{j\in\mathcal{N}_{u_i}} \left(\sum_{k}|M_{[jk]}|\right) \gamma_M^2 +1} , \ \ \ i \in \mathcal{N}
\end{align}

Using~\eqref{eq:pesi}, we can enforce \eqref{eq:paramproof2} by imposing $\gamma_i$ such that
\begin{align}
\label{eq:paramproof3}
    \gamma_i^2 = & \frac{1}{1+ \max_{j\in\mathcal{N}_{p_i}} \left(\sum_{k}|M_{[kj]}|\right) + z_i^2} \nonumber \\
    &  \cdot \frac{\gamma_M^2}{\max_{j\in\mathcal{N}_{u_i}} \left(\sum_{k}|M_{[jk]}|\right) \gamma_M^2 +1}, \ \ \ i \in \mathcal{N}
\end{align}
with $z \in \mathbb{R}$. Finally, we observe that with this choice of $gamma_i$, conditions \eqref{eq:shur2} and \eqref{eq:step1proof2} are always satisfied. Notice that we can use \eqref{eq:paramproof3} to define the mapping $\nu(z)_i:=\gamma_i$ as in \eqref{eq:param}, which is well-defined for every value of $z_i \in \mathbb{R}$. Notice that the mapping $\nu$ essentially provides a free parametrization of the set of all $\gamma_i$ for each sub-operator such that the operator $\Sigma_{\theta}^{M}$ has finite $L_{i2}$-gain $\gamma_{M}$. The result follows by defining the composed mapping $\psi_i(\xi_i, \nu(z_i))$ for each sub-operator.
\end{proof}

Theorem \ref{thm:freeparam} allows us to tackle a learning problem, such as system identification, by solving an unconstrained optimization problem using a distributed operator that takes into account the sub-structure of the model that we want to learn while ensuring an $L_{i2}$-boundedness property of the operator at each step of the optimization process.

\begin{rem}
It is worth mentioning that incorporating the gain $\gamma_{M}$ as an extra free parameter within the proposed parametrization can offer added flexibility when employing this parametrization in a training framework.
\end{rem}

\begin{rem}
Following up Remark \ref{rmk:cons}, we can notice that the free parametrization that we obtained through the mapping $\nu(z)$ is not surjective as we have introduced various degrees of conservatism, for example by using Gershgorin Theorem, which gave us a sufficient but not necessary condition for the positive semi-definiteness of the matrix~\eqref{eq:SchurLMI}. Additional conservatism is present in the parametrization $\psi$ itself. In other words, the free parametrization given by $\psi_i(\xi_i, \nu(z_i))$ allows one to optimize only over a subset of $L_{i2}$-bounded sub-operators that, once interconnected via the matrix $M$, define an $L_{i2}$-bounded distributed operator. As mentioned before, this drawback is greatly outweighed by the advantage of allowing one to solve an unconstrained optimization problem, as shown in the following example.  
\end{rem}

A detailed description of how Theorem~\ref{thm:freeparam} can be leveraged for the training of a network of $L_{i2}$-bounded operators, is provided by Algorithm~\ref{alg:training}. 
\begin{algorithm}
\caption{Training distributed $L_{i2}$-bounded operators}\label{alg:training}
\begin{algorithmic}
\State \textbf{Input:} Training data $(\tilde{x}_0,\tilde{d}, \tilde{y})$, learning rate $\eta$, number of epochs $E$.
\State Initialize randomly the parameters of each subsystem: $\theta_i$, $z_i$ $\forall i\in\mathcal{N}$ and {$\gamma_M$} .

\For{$e=1$ to $E$}
    \State Initialize hidden state $x_i$.
    \State Compute $\gamma_i=\nu(z_i)$ using~\eqref{eq:param}.
    \State Compute sequences $y_{0:T} = \Sigma_{\theta} \left(\tilde{x}_0, \tilde{d}_{0:T} \right)$.
        \State Compute loss $\mathcal{L}(y_{0:T},\tilde{y}_{0:T})$ with~\eqref{eq:loss}.

    \State Backpropagate gradients through time (see~\cite{werbos1990backpropagation}) to update parameters via gradient descent with learning rate $\eta$.
\EndFor
\State \textbf{Output:} Trained model parameters $\theta_i, z_i$, $\forall i\in\mathcal{N}$.
\end{algorithmic}
\end{algorithm}

\section{Numerical example} \label{S:num_example}
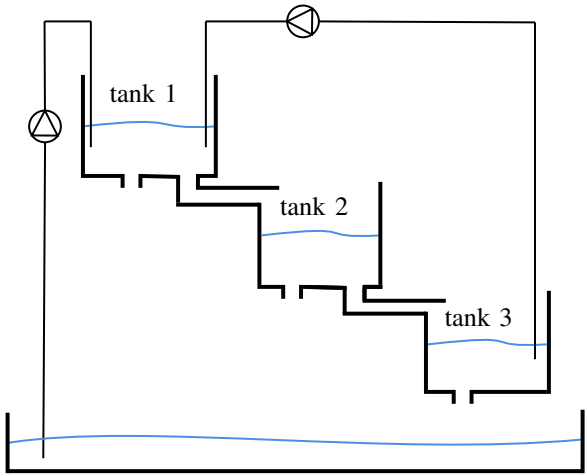
\begin{figure}
    \centering
\tikzset{every picture/.style={line width=0.75pt}} 
\begin{tikzpicture}[x=0.75pt,y=0.75pt,yscale=-1,xscale=1]
\draw [line width=1.5]    (58,101) -- (78,101) ;
\draw [line width=1.5]    (116,101) -- (125,101) ;
\draw [line width=1.5]    (87,101) -- (107,101.5) ;
\draw [line width=1.5]    (87,100) -- (87,107) ;
\draw [line width=1.5]    (78,100) -- (78,107) ;
\draw [color={rgb, 255:red, 74; green, 144; blue, 226 }  ,draw opacity=1 ][line width=0.75]    (58,76) .. controls (111,70) and (89,78) .. (125,76) ;
\draw [line width=1.5]    (116,100) -- (116,107) ;
\draw [line width=1.5]    (106,101) -- (106,114.5) ;
\draw [line width=1.5]    (147,114.5) -- (147,156) ;
\draw [line width=1.5]    (146,157) -- (159,157) ;
\draw [line width=1.5]    (200,157) -- (209,157) ;
\draw [line width=1.5]    (171,157) -- (191,157.5) ;
\draw [line width=1.5]    (168,156) -- (168,163) ;
\draw [line width=1.5]    (159,156) -- (159,163) ;
\draw [color={rgb, 255:red, 74; green, 144; blue, 226 }  ,draw opacity=1 ][line width=0.75]    (148,131) .. controls (201,125) and (172,133) .. (208,131) ;
\draw [line width=1.5]    (200,157) -- (200,164) ;
\draw    (120,23) -- (161,23) ;
\draw   (161,23) .. controls (161,18.58) and (164.58,15) .. (169,15) .. controls (173.42,15) and (177,18.58) .. (177,23) .. controls (177,27.42) and (173.42,31) .. (169,31) .. controls (164.58,31) and (161,27.42) .. (161,23) -- cycle ;
\draw   (161,23) -- (173.67,16.67) -- (173.51,29.64) -- cycle ;
\draw    (120,23) -- (120,86.5) ;
\draw    (177,23) -- (286,23.5) ;
\draw [line width=1.5]    (147,115.5) -- (105,115.5) ;
\draw [line width=1.5]    (157,107) -- (115,107) ;
\draw [line width=1.5]    (208,104) -- (208,156) ;
\draw [line width=1.5]    (200,157) -- (209,157) ;
\draw [line width=1.5]    (169,157) -- (189,157.5) ;
\draw [line width=1.5]    (200,156) -- (200,163) ;
\draw [line width=1.5]    (190,157) -- (190,170.5) ;
\draw [line width=1.5]    (231,169.5) -- (231,211) ;
\draw [line width=1.5]    (293,159) -- (293,211) ;
\draw [line width=1.5]    (232,210) -- (245,210) ;
\draw [line width=1.5]    (254,210) -- (292,210) ;
\draw [line width=1.5]    (254,209) -- (254,216) ;
\draw [line width=1.5]    (245,209) -- (245,216) ;
\draw [color={rgb, 255:red, 74; green, 144; blue, 226 }  ,draw opacity=1 ][line width=0.75]    (232,186) .. controls (285,180) and (256,188) .. (292,186) ;
\draw [line width=1.5]    (231,170.5) -- (189,170.5) ;
\draw [line width=1.5]    (241,164) -- (199,164) ;
\draw    (286,23.5) -- (286,193.5) ;
\draw    (38.64,23) -- (38.83,68) ;
\draw   (38.83,68) .. controls (43.25,67.98) and (46.85,71.55) .. (46.87,75.97) .. controls (46.88,80.39) and (43.32,83.98) .. (38.9,84) .. controls (34.48,84.02) and (30.88,80.45) .. (30.87,76.03) .. controls (30.85,71.62) and (34.41,68.02) .. (38.83,68) -- cycle ;
\draw   (38.83,68) -- (45.22,80.64) -- (32.24,80.54) -- cycle ;
\draw    (38.9,84) -- (38,243.5) ;
\draw    (62,23) -- (38.64,23) ;
\draw    (62,23) -- (62,86.5) ;
\draw [color={rgb, 255:red, 74; green, 144; blue, 226 }  ,draw opacity=1 ][line width=0.75]    (20,235.75) .. controls (87,225) and (215,243) .. (310,234) ;
\draw [line width=1.5]    (20,250) -- (310,250) ;
\draw [line width=1.5]    (310,219) -- (310,251) ;
\draw [line width=1.5]    (20,220.5) -- (20,251) ;
\draw [line width=1.5]    (58,50) -- (58,102) ;
\draw [line width=1.5]    (125,50) -- (125,102) ;
\draw (70,53) node [anchor=north west][inner sep=0.75pt]   [align=left] {tank 1};
\draw (156,110) node [anchor=north west][inner sep=0.75pt]   [align=left] {tank 2};
\draw (239,166) node [anchor=north west][inner sep=0.75pt]   [align=left] {tank 3};
\end{tikzpicture}
    \caption{Triple-tank system with recirculation pump.}
    \label{fig:example}
\end{figure}
\begin{figure}
    \centering
    \includegraphics[width=\columnwidth]{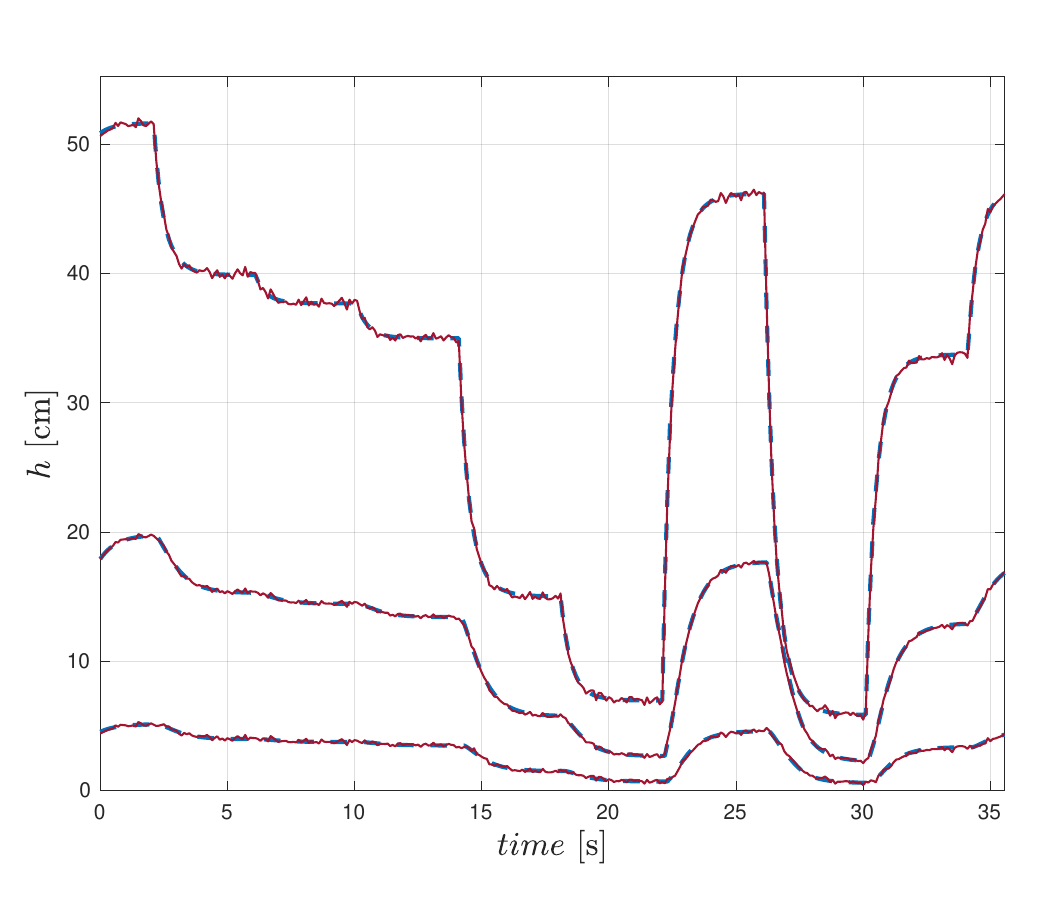}
    \caption{Comparison of the open-loop prediction of the trained distributed RENs built with the proposed parametrization (blue dashed line) versus ground truth (red continuous line) on an independent validation dataset.}
    \label{fig:ouput}
\end{figure}
\begin{figure}
    \centering
    \includegraphics[width=\columnwidth]{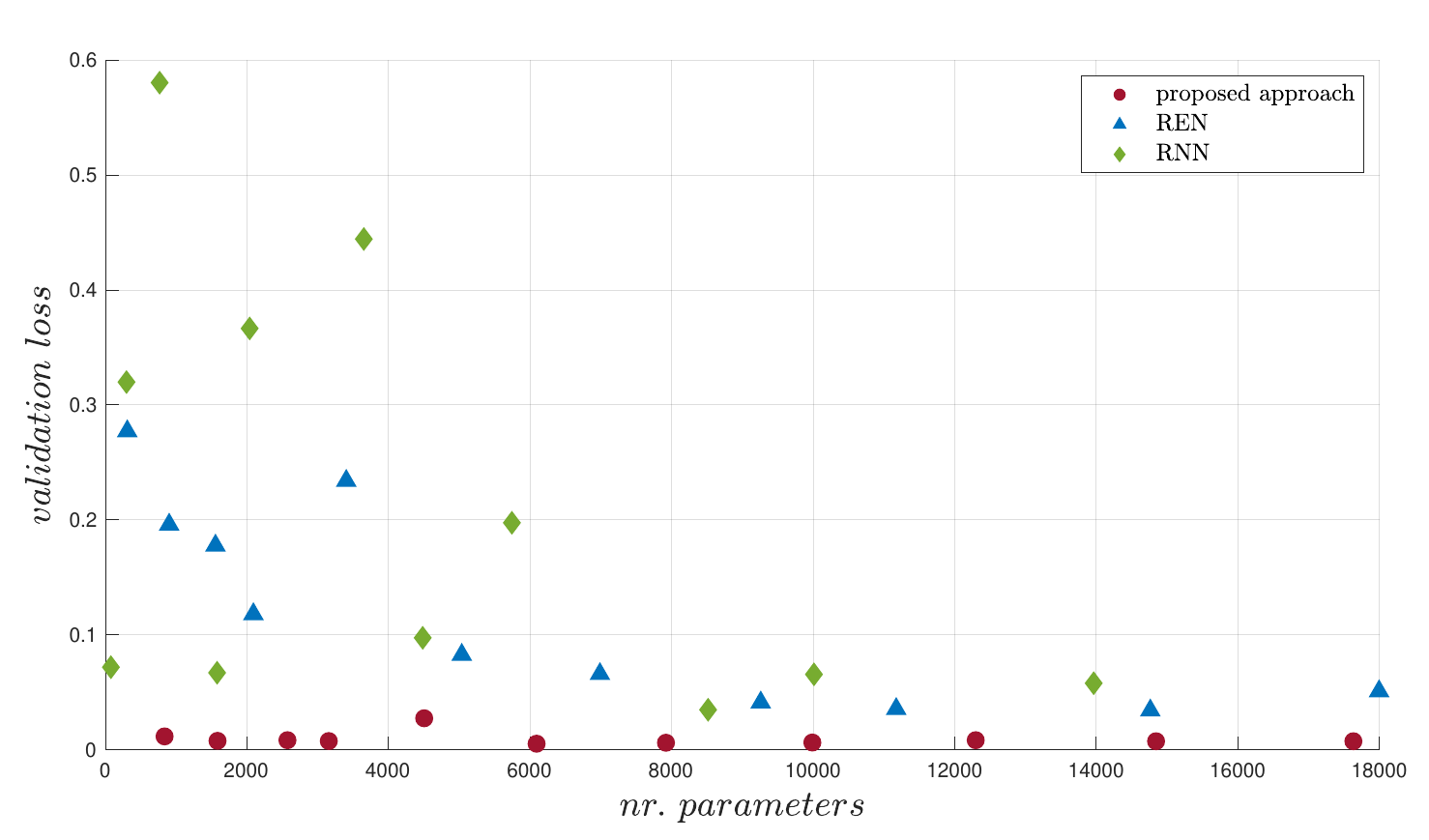}
    \caption{Comparison of validation loss as a function of tunable parameters for the proposed approach with three RENs (red ``$\circ$''), the REN (blue ``$\triangle$'') and the RNN (green ``$\diamond$'').}
    \label{fig:comparison}
\end{figure}

In this section, we validate our results on the identification of the interconnected nonlinear dynamical system shown in Fig.~\ref{fig:example}.
The system is composed of three interconnected tanks and a recirculation pump that allows the water to flow from the third tank to the first. The pump itself cannot be controlled and continuously moves water from the last tank to the first. Additionally, an external controlled pump feeds into the first tank. The first-principles model, describing the dynamics of the system, is 
\begin{align} \label{eq:example}
    \dot{h}_1 &= -\frac{a_1}{A_1}\sqrt{2gh_1}+k_1\frac{a_3}{A_1}\sqrt{2gh_3} + k_c\frac{v}{A_1} \nonumber \\
    \dot{h}_2 &=-\frac{a_2}{A_2}\sqrt{2gh_2}+k_2\frac{a_1}{A_2}\sqrt{2gh_1}\\
    \dot{h}_3 &=-\frac{a_3}{A_3}\sqrt{2gh_3}+k_3\frac{a_2}{A_3}\sqrt{2gh_2} \nonumber
\end{align}
where $A_i$ is the cross-section of the $i$-th tank, $a_i$ is the cross-section of the outlet holes, $k_i$ is the percentage of flow flowing in the next tank, $h_i$ is the water level and $v$ is the inlet flow to tank 1. The parameters employed in the simulation are listed in Tab~\ref{tab:parameters}.
The system~\eqref{eq:example} has been discretized using a Forward Euler scheme with a sampling time $T_s=0.1$. A dataset of $(\tilde{v}, \tilde{h})$ pairs was collected by exciting the system with a random input signal $v$ with values in $[10,\ 100]$. Note that the simulated output measurements have been corrupted by white noise, i.e., $\tilde{h}=h+w$.
The dataset was divided into two subsets: the identification set, which represented $70 \%$ of the entire dataset and was used for training, and the remaining portion of the dataset, which was employed to cross-validate the identified model.
To identify the system~\eqref{eq:example}, we validate the efficacy of the parametrization provided by Theorem~\ref{thm:freeparam} by considering the interconnection of three RENs models~\cite{revay2023recurrent,revay2021recurrent}. Each REN is parametrized following the method described in~\cite{revay2021recurrent}, ensuring that each resulting sub-operator possesses a finite $L_{i2}$-gain and, therefore, belongs to the set $\Phi_{\gamma}$. Employing the proposed approach, we interconnect the RENs as follows:
\begin{equation}
    u= \left[\begin{array}{c}
h_3 \\
v \\
h_1 \\
h_2
\end{array}\right] =
\underbrace{\left[\begin{array}{ccc}
0 & 0 & 1 \\
0 & 0 & 0 \\
1 & 0 & 0 \\
0 & 1 & 0
\end{array}\right]}_{M} \left[\begin{array}{c}
h_1 \\
h_2 \\
h_3 
\end{array}\right] + \underbrace{\left[\begin{array}{c}
0 \\
v \\
0 \\
0
\end{array}\right]}_d.
\end{equation}
To evaluate the efficacy of the proposed approach, we conducted numerical experiments comparing the system identification results obtained using the proposed approach and different neural networks architecture that ignores the topology of the system.
Specifically, we used a Mean Square Error (MSE) as loss function and we compared the proposed approach to a single REN, and a RNN~\cite{medsker2001recurrent}, another dynamical operator with no guarantees regarding $L_2$-boundedness.
Fig.~\ref{fig:comparison} shows the obtained loss on the independent validation dataset as a function of the number of tunable parameters in the considered models. For the comparison, all three architectures were implemented in PyTorch with a learning rate of $\eta = 10^{-2}$ and trained for $E=500$ epochs. The implementation of the proposed approach can be found at: \url{https://github.com/DecodEPFL/DistributedREN}.
As evident from Figure~\ref{fig:comparison}, the proposed architecture, leveraging prior knowledge of the system topology, achieves significantly improved performance in terms of prediction error on the validation dataset with a notably reduced number of tunable parameters, which are essentially optimization variables in the optimization problem. As expected all models exhibit better performance as the number of tunable parameters increases.
Fig.~\ref{fig:ouput} shows the predicted states of the identified model in blue and the trajectories of the validation dataset in red.

\begin{table}
\vspace{2mm}
\begin{center}
\begin{tabular}{ | c | c | c | c| c| c| } 
  \hline
 $A_1$& $38 \ cm^2$ & $A_2$ & $ 32 \ cm^2$ \\ 
  \hline
  $A_3$& $ 21 \ cm^2$ & $a_1$ & $0.05 \ cm^2$ \\ 
  \hline
  $a_2$& $0.03 \ cm^2$ & $a_3$ & $0.06 \ cm^2$ \\ 
  \hline
  $k_1$& $0.32 $ & $k_2$ & $0.23$ \\ 
  \hline
  $k_3$& $0.52$ & $k_c$ & $50$ \\ 
  \hline
\end{tabular} \caption{List of parameters employed in the simulation.}
\label{tab:parameters}
\end{center}
\end{table}


\section{Conclusions}
We have introduced a novel free parametrization for incremental $L_2$ operators for the identification of distributed systems by leveraging prior knowledge of system behavior and sparsity pattern. A natural extension of the approach would involve exploring additional forms of dissipativity for interconnected systems. Moreover, the potential application of these operators as distributed controllers, mirroring the sparsity pattern of the plant, holds promise for enhancing closed-loop properties based on dissipativity theory. The versatility and potential impact of this methodology offer exciting prospects for the continued development and refinement of identification and design control strategies in complex distributed nonlinear systems.

\bibliographystyle{IEEEtran}
\bibliography{Main_v1}    

\end{document}